\documentclass{moriond}

\def\Journal#1#2#3#4{{#1} {\bf #2} #3 (#4)}

\def\be{\begin{equation}}
\def\ee{\end{equation}}
\def\bea{\begin{eqnarray}}
\def\eea{\end{eqnarray}}

\newcommand{\Photo}{}

\begin{document}
\vspace*{4cm}
\title{Jet substructure in pp collisions with ALICE}

\author{ Ezra D. Lesser, on behalf of the ALICE Collaboration\\ }

\address{UC Berkeley, Department of Physics, 366 Physics North, Berkeley, CA 94720-7300, USA}

\maketitle\abstracts{
Jet substructure observables have been used by experiments at the Large Hadron
Collider (LHC) as instruments to search for new physics as well as to study
perturbative and nonperturbative processes in quantum chromodynamics (QCD).
Some observables are infrared and collinear safe and thus easily comparable
to first-principles calculations, while others offer direct insight into specific
physical phenomena such as the quark-gluon plasma formed in heavy-ion collisions.
The high-precision capability of the ALICE tracking system allows a unique
opportunity at LHC energies to measure tracks with low transverse momentum,
permitting high precision access to the softer components inside jets with an
excellent angular resolution. We present some recent charged-particle jet
substructure measurements in pp collisions with ALICE, including the generalized
jet angularities, the angular jet axes differences, and the direct observation of
the dead-cone effect in QCD using the primary Lund Plane. These results provide
new insights into the evolution of jets by comparing ALICE measurements to
predictions from different event generators and perturbative QCD calculations.}

%%%%%%%%%%%%%%%%%%%%%%%%%%%%%%%%%%%%%%%%%%%%%%%%%%%%%%%%%%%%%%%%%%%%%%%%%%
\section{Introduction}

In high-energy particle collisions, hadron jets are a direct consequence
of quantum chromodynamics (QCD). These jets form from an initial hard
(high $Q^2$) parton scattering, followed by a scale evolution culminating in
hadronization near $\Lambda_\mathrm{QCD}$. Jets can be experimentally
defined from a set of measured tracks by using a jet reconstruction algorithm,
such as a sequential recombination algorithm, along with a chosen resolution
(radius) parameter. %, which modifies sensitivity to nonperturbative effects.
Jet substructure observables are then calculated from the jet's individual
constituents, allowing characterization
of the jet's internal radiation pattern. Some observables are constructed
to enable direct calculations from first-principles QCD,
permitting direct tests of theory. Jet substructure
observables can additionally be tuned to explore nonperturbative processes
such as hadronization.

Jet ``grooming'' procedures can be used to remove soft,
wide-angle radiations and enhance perturbative calculability. One of the
most popular grooming procedures is known as soft drop~\cite{soft_drop},
which first reclusters jet constituents using the Cambridge-Aachen
algorithm~\cite{CA_algorithm}, producing a tree data structure that
follows the expected angular ordering of emissions in QCD. One can then
iterate through this tree, trimming away the outermost branches until the
soft drop grooming condition,
$\min(p_\mathrm{T1}$, $p_\mathrm{T2})/(p_\mathrm{T1} + p_\mathrm{T2}) >
z_\mathrm{cut} \left( \Delta R_{12}/R \right)^\beta$, is satisfied,
where $R$ is the jet radius, $p_{T1}$ and $p_{T2}$ are the transverse momenta
of two branches at a particular vertex, $\Delta R_{12}$ is their distance in
the rapidity-azimuth plane, and $z_\mathrm{cut}$ and $\beta$ are
user-defined parameters.

In these proceedings, each of the following three sections highlights a recent
measurement of jet substructure carried out by ALICE using data recorded at the
LHC from pp collisions at $\sqrt{s} = 5.02$ or 13 TeV.
In particular, Sec.~\ref{sec:ang} reviews the recent ALICE measurements of
the jet angularities,
Sec.~\ref{sec:axis} shows recent ALICE measurements of the jet axis differences,
and Sec.~\ref{sec:deadcone} discusses the first-ever direct measurement of the
dead-cone effect in QCD. Jets were reconstructed using charged-particle tracks
and the anti-$k_T$ algorithm~\cite{anti-k_T} with $E$-scheme recombination.

%%%%%%%%%%%%%%%%%%%%%%%%%%%%%%%%%%%%%%%%%%%%%%%%%%%%%%%%%%%%%%%%%%%%%%%%%%
\section{Groomed and ungroomed jet angularities}
\label{sec:ang}

One interesting set of jet substructure observables is the generalized jet
angularities $\lambda_{\alpha}^{\kappa}$, with
\begin{equation} \label{eqn:jet_ang}
\lambda_{\alpha}^{\kappa}
\equiv \sum_i \left( \frac{p_{\mathrm{T},i}}{p_{\mathrm{T,jet}}} \right)^\kappa
              \left( \frac{\Delta R_i}{R} \right)^\alpha
\equiv \sum_i z_i^{\kappa} \theta_i^{\alpha},
\end{equation}
where $i$ runs over the jet constituents, $R$ is the jet radius,
$\Delta R_i$ is the distance from constituent $i$ to the jet axis in
the rapidity-azimuth plane, and $\kappa$ and $\alpha$ are continuous,
tunable parameters. These observables are infrared- and collinear- (IRC-)safe
when $\kappa=1$ and $\alpha > 0$, allowing direct calculations from
perturbative QCD (pQCD). By varying the angular weighting with $\alpha$
as well as the fragmentation bias with $R$, one can probe different regions
of emission phase space.

ALICE has recently measured~\cite{alice_ang} the jet angularities
$\lambda_\alpha^{\kappa=1}$ both with and without soft drop grooming
for $R \in \{0.2$, $0.4\}$ and $\alpha \in \{1$, 1.5, 2, $3\}$.
Agreement within about 40\% is observed with respect to
PYTHIA8 Monash 2013~\cite{pythia_manual} and
Herwig 7 (default tune)~\cite{herwig_manual} Monte Carlo (MC) generators,
with noted improvement in the groomed with respect to the ungroomed case.
This reflects upon the MC parton shower and fragmentation models, which
improve when nonpertubative effects, under less stringent control in the
phenomenological approach, are reduced in the grooming step.

\begin{figure}[!b]
    \centering
    \includegraphics[scale=0.7]{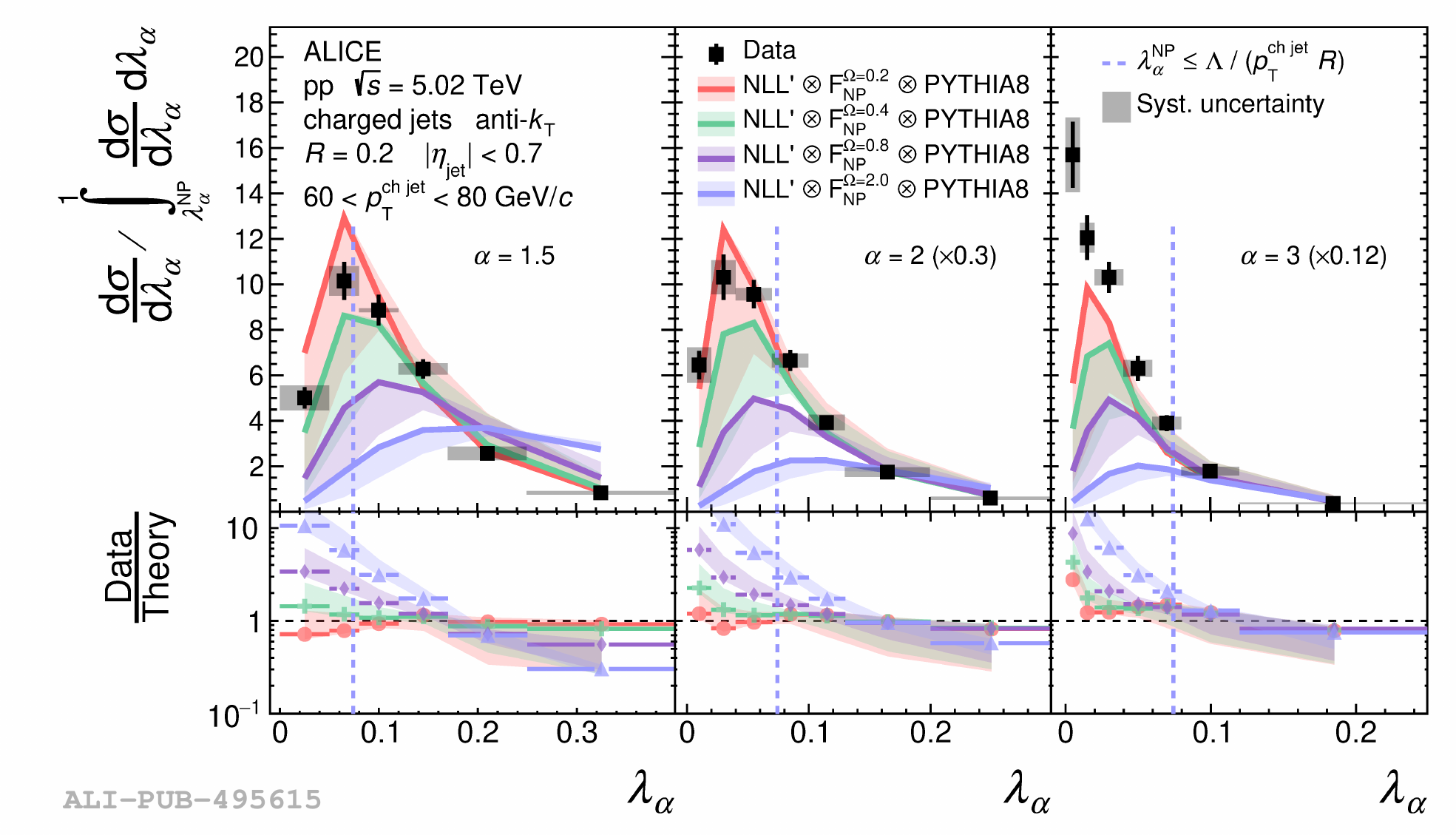}
    \caption{Example comparison of the ungroomed jet angularities with different values of $\alpha \in \{1.5$, 2, $3\}$ to NLL$^\prime$ SCET calculations convolved with different amounts of smearing with a nonperturbative shape function~\protect\cite{alice_ang}. Distributions are normalized to the perturbative region, right of the vertical dashed line, which uses $\Lambda=1$ GeV.}
    \label{fig:ang_theory}
\end{figure}

Comparisons were also carried out between ALICE data and predictions from
Soft-Collinear Effective Theory (SCET)~\cite{scet_ang,scet_ang_gr}.
Nonperturbative corrections were handled in two separate and independent ways.
One method was convolution with a
nonperturbative shape function~\cite{shape_fn} $F(k)$,
\begin{equation}
F(k) = \frac{4k}{\Omega_\alpha^2} \exp \left( -\frac{2k}{\Omega_\alpha} \right)
\hspace{1em} \mathrm{with} \hspace{1em}
\frac{\mathrm{d}\sigma}{\mathrm{d}p_\mathrm{T}\; \mathrm{d}\lambda_\alpha} = 
\int \mathrm{d}k \; F(k) \;
\frac{\mathrm{d}\sigma^\mathrm{pert}}{\mathrm{d}p_\mathrm{T}\; \mathrm{d}\lambda_\alpha}
\left( \lambda_\alpha - \frac{k}{p_\mathrm{T} R} \right),
\end{equation}
where $k$ is an integration parameter with units of momentum,
$p_\mathrm{T}$ is the jet transverse momentum,
$\Omega_\alpha = \Omega / (\alpha - 1)$, and $\Omega$ is an unknown but
presumably universal parameter with units of momentum.
This smearing is also followed by a folding to correct for charged-particle
jet effects. Figure~\ref{fig:ang_theory} shows comparisons to theoretical
predictions with four different values $\Omega$. Smaller values on the order of
$0.2-0.4$ GeV/$c$ give the best agreement to the measured spectra.
As $\alpha$ is increased and the wider-angle constituents are correspondingly
more strongly emphasized, the curves are pushed into the nonperturbative region,
and divergence is observed at low $\lambda_\alpha$, while
the perturbative region remains relatively well-described.

%%%%%%%%%%%%%%%%%%%%%%%%%%%%%%%%%%%%%%%%%%%%%%%%%%%%%%%%%%%%%%%%%%%%%%%%%%
\section{Jet axis differences}
\label{sec:axis}

The axis of a jet can be defined in different ways. A standard technique is
to use $E$-scheme recombination, where the 4-vectors of each constituent are
summed to obtain the jet 4-vector. One could also follow this
approach only using the constituents that survive soft drop grooming.
Yet another way is the Winner-Take-All (WTA) method, where the jet is first
reclustered into a tree data structure using the Cambridge-Aachen algorithm
and then traversed through following the hardest splitting, with the final
constituent defining the jet's axis. Measuring the difference between these
different jet axis definitions is an IRC-safe way to observe the influence of
soft, nonperturbative radiation, and the observable is also sensitive to
PDFs and TMDs~\cite{R_axis_TMD}.

Measurements of the jet axis differences, $\Delta R_\mathrm{axis}$, have been
carried out by ALICE using the three different definitions above with varying
grooming settings.
The distributions reveal that the jet axis differences between the
standard and groomed axes are strongly correlated, with the distributions
spanning a range much smaller than $R$. The WTA axis, however, tends
to be less strongly aligned with the standard or groomed axes, suggesting that
$p_\mathrm{T}$ tends to be more broadly distributed in the jet, rather than
collimated along a single axis. 

\begin{figure}[!h]
    \centering
    \includegraphics[scale=0.7]{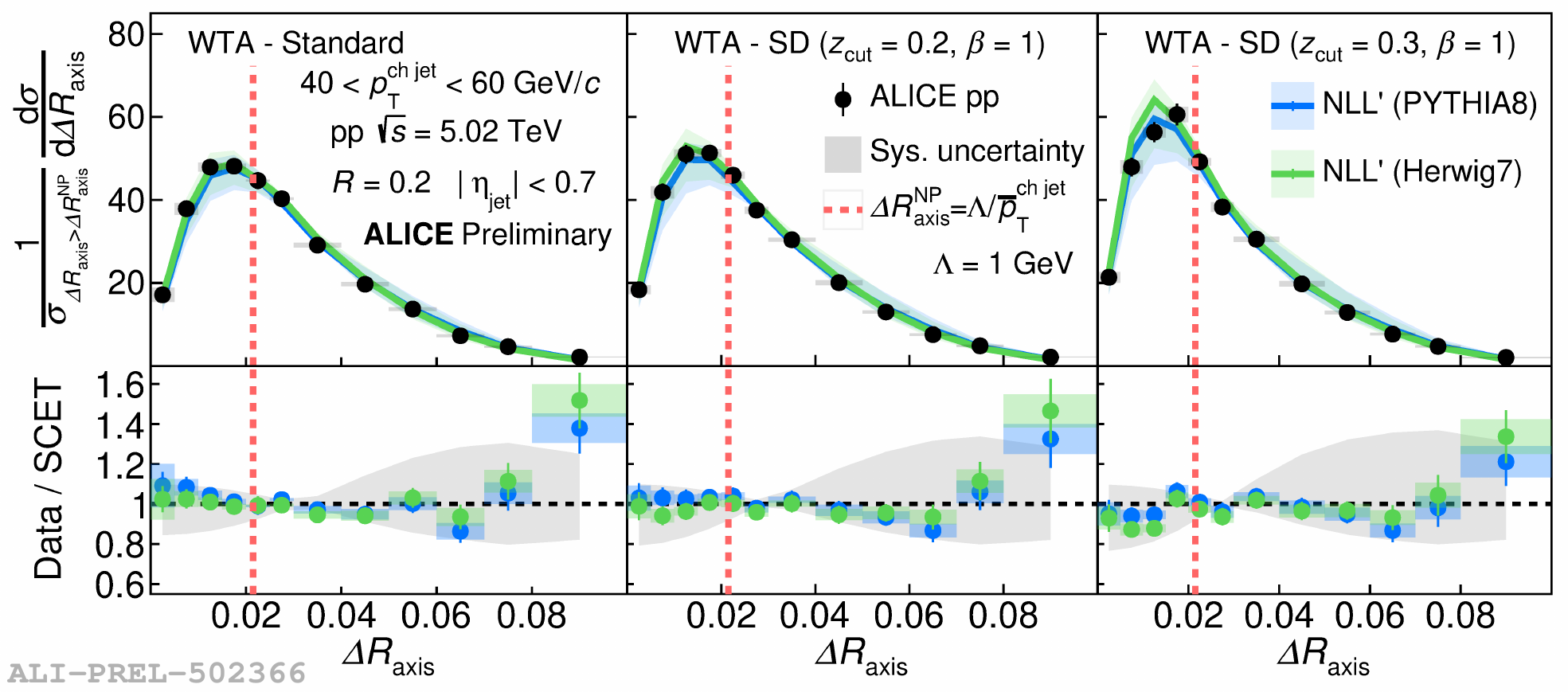}
    \caption{Example comparison of the WTA with standard and soft drop axes. Theory curves are normalized to the perturbative region, right of the vertical dashed line. Good agreement is observed between data and NLL$^\prime$ calculations in both regions, demonstrating good behavior of the calculations within the given uncertainties.}
    \label{fig:axis_theory}
\end{figure}

PYTHIA8 and Herwig 7 are largely able to reproduce these trends, with agreement
observed within about 20\% for these observables. MC predictions for the
WTA differences more closely match ALICE data than the standard-to-groomed ones.
Comparisons to NLL$^\prime$ predictions, using charged-particle jet corrections
from either MC, also show excellent agreement within uncertainties;
an example is given in Fig.~\ref{fig:axis_theory}.

%%%%%%%%%%%%%%%%%%%%%%%%%%%%%%%%%%%%%%%%%%%%%%%%%%%%%%%%%%%%%%%%%%%%%%%%%%
\section{Dead-cone effect in QCD}
\label{sec:deadcone}

The dead-cone effect in QCD is the expected suppression of gluon radiation
within an angle $m/E$ from an emitting parton~\cite{deadcone}.
Effects can be observed by comparing jets initiated by a heavy-flavor quark
with respect to $u$-, $d$-, and $g$-dominated inclusive jets.
A direct measurement is challenged, however, by the need to
identify gluon radiation separately from secondary effects and to
determine the dynamic direction of the quark throughout its parton shower.

\begin{figure}[!h]
    \centering
    \includegraphics[scale=0.7]{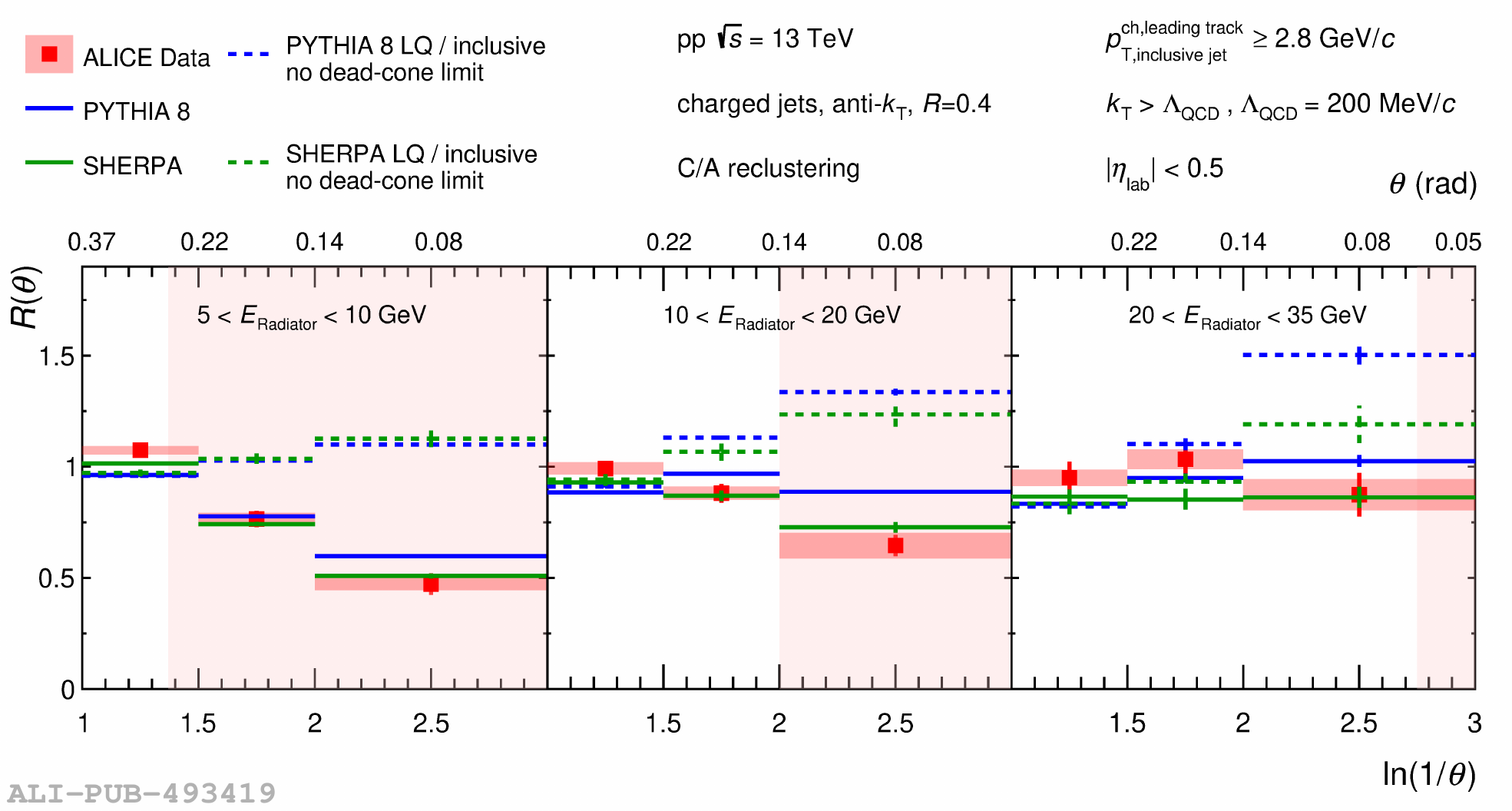}
    \caption{First direct observation of the dead-cone effect in QCD~\protect\cite{alice_deadcone} using $D^0$ versus inclusive jets in different bins of $E_\mathrm{radiator}$. Small angle suppression corresponds to a dip in the shaded red regions at larger $\ln(1/\theta)$.}
    \label{fig:deadcone}
\end{figure}

ALICE has recently performed the first direct measurement of this
effect~\cite{alice_deadcone} by employing the Cambridge-Aachen algorithm to
reconstruct the primary Lund plane for both inclusive and $D^0$-tagged jets,
and then projecting onto the angular axis and taking the ratio between them,
\begin{equation}
R(\theta) = \frac{1}{N^\mathrm{D^0\;jets}}
\frac{\mathrm{d}n^\mathrm{D^0\;jets}}{\mathrm{d} \ln(1/\theta)} \bigg/
\frac{1}{N^\mathrm{inclusive\;jets}}
\frac{\mathrm{d}n^\mathrm{inclusive\;jets}}{\mathrm{d} \ln(1/\theta)}
\bigg|_{k_\mathrm{T},E_\mathrm{radiator}}.
\end{equation}
When compared to the light-quark / inclusive limit, significant suppression
is seen at small angles, with the suppression also decreasing with
$E_\mathrm{radiator}$. This effect is also seen in PYTHIA8 and SHERPA MC
generators, which is expected. Figure~\ref{fig:deadcone} shows the ALICE
measurement.

%%%%%%%%%%%%%%%%%%%%%%%%%%%%%%%%%%%%%%%%%%%%%%%%%%%%%%%%%%%%%%%%%%%%%%%%%%

\end{document}